\documentclass{article}
\usepackage{moreverb}
\usepackage{tikz}
\usepackage[numbers]{natbib}
\usepackage{booktabs}
\usepackage{amsmath}
\usepackage{amsfonts}
\usepackage{algorithm}
\usetikzlibrary{arrows.meta}
\usepackage{graphicx} 
\usetikzlibrary{shapes,positioning}
 \usepackage{url}
\usepackage{moreverb}
\pdfoutput=1 
\newcommand\BibTeX{{\rmfamily B\kern-.05em \textsc{i\kern-.025em b}\kern-.08em
T\kern-.1667em\lower.7ex\hbox{E}\kern-.125emX}}

\usepackage{subcaption}
\usepackage{enumitem}
\usepackage{comment}

\newcommand{\pkg}[1]{{\normalfont\fontseries{b}\selectfont #1}}  
\newcommand{\indep}{\perp \!\!\! \perp}
\setlist[description]{font=\normalfont\itshape\textbullet\space}

\newcommand{\ks}[1]{\textcolor{black}{#1}}

\begin{document}

\title{WATCH: A Workflow to Assess Treatment Effect Heterogeneity in\\ Drug Development for Clinical Trial Sponsors}

\date{}
\author{Konstantinos Sechidis\thanks{Advanced Methodology and Data Science, Novartis Pharma AG, Basel, Switzerland}
\and Sophie Sun\thanks{Advanced Methodology and Data Science, Novartis Pharmaceuticals Corporation, East Hanover, New Jersey, USA}
\and Yao Chen\footnotemark[2]
\and Jiarui Lu\footnotemark[2]
\and Cong Zhang\thanks{China Novartis Institutes for Bio-medical Research CO., Shanghai, China}
\and Mark Baillie\footnotemark[1]
\and David Ohlssen\footnotemark[2]
\and Marc Vandemeulebroecke\thanks{Development Analytics and Statistical Innovation, UCB Farchim SA, Bulle, Switzerland}
\and Rob Hemmings\thanks{Consilium Hemmings (UK) Ltd., Woking, UK}
\and Stephen Ruberg\thanks{Analytix Thinking, LLC, Indianapolis, Indiana, USA}
\and Björn Bornkamp\footnotemark[1]
}



\maketitle
\abstract{This paper proposes a Workflow for Assessing Treatment effeCt Heterogeneity (WATCH) in clinical drug development targeted at clinical trial sponsors. WATCH is designed to address the challenges of investigating treatment effect heterogeneity (TEH) in randomized clinical trials, where sample size and multiplicity limit the reliability of findings. The proposed workflow includes four steps: Analysis Planning, Initial Data Analysis and Analysis Dataset Creation, TEH Exploration, and Multidisciplinary Assessment. \ks{The workflow offers a general overview of how treatment effects vary by baseline covariates in the observed data, and guides interpretation of the observed findings based on external evidence and best scientific understanding. The workflow is exploratory and not inferential/confirmatory in nature, but should be pre-planned before data-base lock and analysis start. It is focused on providing a general overview rather than a single specific finding or subgroup with differential effect.}



\maketitle

\renewcommand\thefootnote{}

\renewcommand\thefootnote{\fnsymbol{footnote}}
\setcounter{footnote}{1}
\section{Introduction}
\label{sec:intro}
Clinical trial sponsors make important decisions in drug development based on internal exploratory analyses, for example related to treatment effect heterogeneity (TEH), which refers to how treatment effects may vary according to patient baseline variables or patient subgroups\footnote{In the clinical trials literature and drug development, investigation of treatment effects in patient subgroups is often called \emph{subgroup analysis}.}.
Accounting for treatment effect heterogeneity is an important task for sponsors in drug development, impacting, for example, the selection of inclusion/exclusion criteria of planned clinical trials, or whether to conduct follow-up trials in a more targeted population when an earlier trial showed limited efficacy in the overall population. These analyses are also important, even when no further concrete studies are planned, to thoroughly understand the available data for any trends in treatment effect heterogeneity. Correspondingly investigation of treatment effect heterogeneity is a crucial task during drug development.

On the other hand, it is well known that statements around treatment effects in subgroups within a single clinical trial (or a small number of clinical trials) are unreliable: There is considerable empirical evidence that findings around treatment effects in subgroups rarely get replicated\citep{Yusuf1991,wallach2017evaluation}. The reason is simple: the sample size for clinical trials is generally determined to be able to demonstrate a treatment effect for the overall trial population. A sample size that is sufficient for this purpose, will not be sufficient to make definitive statements on either (i) the existence of a treatment effect in a subgroup (which will have a reduced sample size), or (ii) the difference in treatment effects across a subgroup and its complement (interaction tests, see also Gelman et al.\cite[Ch. 16]{gelman2020regression}). Furthermore, assessing treatment effects across various baseline covariates or patient subgroups induces a substantial multiplicity problem: When focusing on the subgroups with best or worst observed treatment effect, estimates will suffer \ks{from random highs (or random lows), in particular when focusing on small subgroups\citep{bretz2014multiplicity}.}

Due to these complexities, which naturally arise in all randomized clinical trials of any size and duration, this has sometimes been called \emph{the hardest problem there is}\citep{ruberg2021assessing}. Despite these statistical problems, the regulators make clear that learning how treatment effects can vary between baseline patient variables provides important information. For example EMA\citep{ema:2019}, in its exploratory subgroup guidance for confirmatory trials stresses the regulatory importance of investigating treatment effects according to important subgroups. Similarly, Amatya et al.\citep{amatya2021subgroup} illustrate, based on oncology examples, the vital role of subgroup analyses in the decision making of the {FDA}, which, in specific cases, may lead to label restrictions despite an overall positive trial, or extension of the label to the entire study population despite positive results appearing primarily in a subgroup. In addition, Hemmings\citep{hemmings2014overview} argues that ``... Pharmacology, biology, and clinical practice are complex, and it may be argued that an assumption of complete homogeneity is rarely credible,'' while, when simply assuming homogeneity there is also an error to make  `` ... without full exploration of subgroups the other potential error, failing to identify a truly different effect in a subgroup of patients, will be made wherever the phenomenon exists. Ignoring the problem, and similarly routinely dismissing results of subgroup analysis, is no scientific solution. ...'' 

In terms of communication Amatya et al.\citep{amatya2021subgroup} (see also Alosh et al.\citep{alosh2017tutorial}) suggest, from a regulatory perspective, a categorization of subgroup analyses according to their reliability for regulatory decision-making. When a subgroup analysis is pre-specified and part of the primary multiple test strategy, it is referred to as \textit{inferential} subgroup analysis. The issues mentioned in the previous paragraphs do not apply to this setting, as the subgroup analysis is appropriately taken into account for study design (by appropriate sample size calculation) and analysis (via appropriate multiplicity adjustment). \ks{The terms \textit{supportive} and \textit{exploratory} subgroup analyses are used to describe analyses that examine the consistency of treatment effects across subgroups in a clinical trial, aiming to gain further insight into potentially predictive mechanistic variables and biological characteristics of the disease, and to generate hypotheses that need to be confirmed in future clinical trials.} The EMA\citep{ema:2019} guideline provides a similar categorization: It defers inferential subgroup problems to the multiplicity guideline, while only exploratory subgroup analyses are in scope for the subgroup guidance.

\subsection*{Our contribution}
We believe the conundrum of, on the one hand, great interest and need, and on the other hand, inherent data limitations (due to sample size and multiplicity) cannot be resolved \ks{solely} by developing a new data analysis methodology but needs to be resolved by (i) following good statistical and data science practices (analysis pre-planning, standardized approaches for data pre-processing,..., see also Baillie et al.\citep{baillie2023good}) (ii) considering external evidence (which may be scarce) and best scientific understanding and (iii) appropriate communication. 

In this article, we propose a Workflow for Assessing Treatment effeCt Heterogeneity (WATCH) for clinical trial sponsors in the \textit{exploratory} setting. Conducting clinical trials is costly and resource intensive. It is a responsibility of the sponsor to learn about, understand and appropriately interpret the generated clinical trial data. As discussed, sponsors can make important internal decisions based on the interpretation of treatment effect heterogeneity in the data: Observed Phase 2 trial results may for example impact how the Phase 3 program is planned, or observed Phase 3 trial results may impact the development program of upcoming compounds with similar mechanisms. A trial may have shown limited efficacy in the overall population so a company may consider running follow-up trials in a more targeted population. These decisions are usually based on internal exploratory analyses.

\ks{We focus on the exploratory setting also because} we think this is where the need is largest. While most of the biostatistical literature provides guidance and tools for formal probability-based statistical inference, this is of limited use for the exploratory setting\citep{tong2019statistical}: To be valid probability based statistical inference requires pre-specification of the model or at least the model selection procedure. But in the exploratory setting, contrary to the confirmatory setting, the model may often be selected in a data-driven manner and iteratively refined based on scientific input and external information, invalidating the assumptions underlying formal statistical inference procedures. Presenting the results of statistical inference procedures, as if the iterative data-driven process had not happened, will be misleading. 

\ks{The past years saw several developments in post-selection inference \cite{sechidis2021,mueller2023isotonicsubgroupselection}. However, they are challenging to use in the exploratory setting that we consider here. It is unrealistic that an exploratory analysis would be done without further refinements such as adding/removing variables and considering further endpoints based on observed results. Adapting and re-running analyses and claiming ``control of a type-I error'' will overstate the conclusions and not be adequate.}


While the regulatory agencies provide clear guidance on pre-specified inferential analyses, we believe there is a gap in terms of exploratory analyses. The EMA guideline on subgroups\citep{ema:2019} provides helpful considerations, but is primarily aimed at assessors in European regulatory agencies and health authority interactions, while we consider the drug developer perspective and primarily sponsor internal decision making. WATCH provides a general overview of how treatment effects may vary by baseline covariates and guides interpretation of the observed findings based on external evidence and best scientific understanding. The workflow is exploratory and not inferential/confirmatory in nature, but should be pre-planned before data-base lock and analysis start. It is focused on providing a general overview rather than a single specific finding or subgroup with differential effect.

\subsection*{Relevant works}
In recent years a number of workflows have been proposed to assess potential treatment effect heterogeneity from very different perspectives and with different aims and purposes. Although the objectives of most of these works are different from ours, we review them here to provide an overview of the current literature. For a comprehensive overview of earlier literature, see Dmitrienko et al.\citep{dmitrienko:2016}. Ruberg and Shen\citep{rube:shen:2015} propose the disciplined subgroup search (DSS) approach (see also Lipkovich et al.\citep{lipk:dmit:agos:2017}). They consider a fully prospective and pre-specified subgroup search algorithm that is intended to make inferential statements about treatment effects in selected subgroups, as well as derive treatment effect estimates. 

Muysers et al.\citep{muysers2020systematic} and Watson and Holmes\citep{watson2020machine} are closest to our proposal. Muysers et al.\citep{muysers2020systematic} focus on gaining an \emph{overall} understanding of how treatment effects vary across many different subgroups. The main proposed tool is a non-inferential, comprehensive, interactive plot showing subgroup treatment effects, which guides a discussion with subject matter experts around the plausibility of findings. This work is exploratory in nature and quite similar in perspective and aims to the workflow we propose in this paper. Watson and Holmes\citep{watson2020machine} emphasize the importance of analysis plans and pre-definition for the investigation of treatment effect heterogeneity. Based on the employed machine learning models, they propose to use a global statistical test to assess either the existence of cross-over interactions (also called qualitative interactions) or whether a more flexible machine learning model outperforms a simpler model without treatment by covariate interactions (similar to \ks{the overall test for the presence of treatment effect heterogeneity}, which we will also adopt in our workflow). However no guidance on how to present the results to stakeholders is provided. 

The Predictive Approaches to Treatment Effect Heterogeneity (PATH) statement was developed and released by a clinical trialist expert panel\citep{kent2020a}. The statement emphasizes the use of \emph{risk modelling}. In this approach, a risk (prognostic) score for the outcome is first derived for each patient. The treatment effect is then assessed in dependence on this risk score using a univariate interaction test. The argument for assessing the treatment effect by the baseline risk score is that it may often be the most important predictor of treatment effect. Schandelmaier et al.\citep{Schandelmaier:2020} recently introduced the \emph{Instrument to assess the Credibility of Effect Modification Analyses} (ICEMAN) in randomized controlled trials and meta-analyses, which is a tool to make post-hoc judgments about the quality of a published subgroup finding, rather than guidance on how to perform the analysis. The following main criteria were identified as most important: (1) Was the direction of the effect modification correctly hypothesized a priori? (2) Was the effect modification supported by prior evidence? (3) Does a test for interaction suggest that chance is an unlikely explanation of the apparent effect modification? (4) Did the authors test only a small number of effect modifiers or consider the number in their statistical analysis? (5) If the effect modifier is a continuous variable, were arbitrary cut-off points avoided? 

\ks{In Section \ref{sec:workflow}, we will present our proposed workflow to assess treatment effect heterogeneity in the exploratory setting and illustrate it with a numerical example. Finally, Section \ref{sec:concl} concludes the paper.}

\section{Methods and Example}
\label{sec:workflow}

Motivated by the established Problem-Plan-Data-Analysis-Conclusion (PPDAC) approach \citep{wild1999statistical, spiegelhalter2019art}, we propose a workflow for assessment of treatment effect heterogeneity in Figure \ref{fig:workflow}. In our case the considered problem is that of assessing treatment effect heterogeneity, and the workflow includes four steps, corresponding to plan, data, analysis, conclusion, but re-framed for our problem: (1) Analysis Planning, (2) Initial Data Analysis and Analysis Dataset Creation, (3) TEH Exploration, and (4) Multidisciplinary Assessment. Step 1 should ideally take place at the planning stage of the trial and entails clarifying the question of interest and deciding on the baseline covariates and outcome(s) to include based on existing knowledge. In step 2, an initial data analysis and analysis data-set creation are more operational tasks, but they can have an important impact on final results. Step 3 is the core analytical step, which is broken down into 3 sub-steps. \ks{The first focus is on an overall (test) assessment of evidence against homogeneity, as proposed in some of the workflows mentioned above.} Then it is investigated which patient baseline variables are associated with the treatment effect, and finally graphical displays are shown to describe the treatment effect heterogeneity. The overall test adjusts for all variables included (i.e. takes into account multiplicity) and serves as a \emph{strength of evidence} assessment for the remaining analytical steps: In case of low evidence against homogeneity the risk of \ks{random highs or random lows} is high and subsequent results should be cautiously interpreted. Finally, the findings need to be assessed in a multidisciplinary team review in step 4.

One may wonder why in step 3 subgroup identification is not explicitly mentioned. We consider the workflow to be a first step after trial read-out and thus focusing on providing a general overview on how the treatment effect may vary according to baseline covariates, but not necessarily to identify a concrete subgroup. A methodology that identifies only concrete subgroups may not adequately provide a global overview of treatment effect heterogeneity (see also the discussion in Bornkamp et al.\citep{bornkamp2024predicting}). \ks{Instead, WATCH assessesment of treatment effect heterogeneity is less inferential in nature, but possibly more suitable for the exploratory setting we consider.} We will return to this in Section \ref{sec:cfa}.

\begin{figure}[h!]
  \begin{center}
    \includegraphics[width=0.4\textwidth]{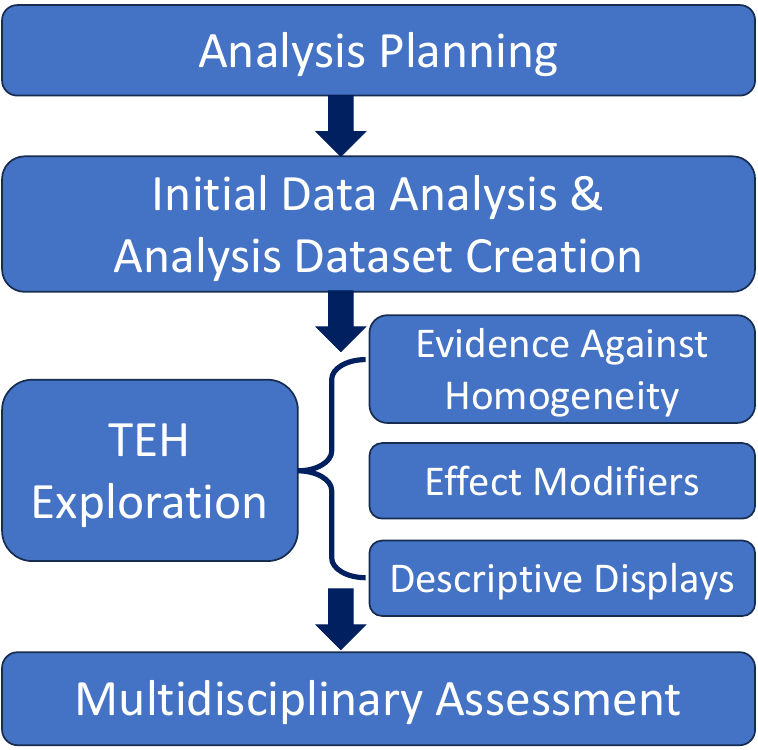}
  \end{center}
  \caption{Overview of WATCH workflow and the four main steps: (1) Analysis Planning, (2) Initial Data Analysis and Analysis Dataset Creation, (3) TEH Exploration, and (4) Multidisciplinary Assessment.}
  \label{fig:workflow}
\end{figure}

In the following sections we will first present the steps above in more detail. To illustrate potential graphical and numerical outputs, a simulated dataset is used from \pkg{benchtm} R package \citep{benchtm} to mimic clinical trial data.  Sun et al.\citep{sun2024comparing} provides details on how the different scenarios are simulated. \ks{Specifically in our demonstration example a random sample with sample size 500 and 30 baseline covariates $X_1, X_2, \ldots, X_{30}$ are generated, with random treatment assignment $P(A = 1) = P(A = 0) = 0.5$. The outcome is generated from $Y \sim N(\mu, \sigma = 1)$ where $\mu = 1.38*((X_1 = 'N') - 0.5*X_{17}) + A*(-0.105 + 0.725*(X_{14} > 0.25)*(X_1 = 'N'))$. Covariates $X_1, X_2, \ldots, X_{30}$ are synthetic data generated from real clinical trial maintaining same covariates structure (see Sun et al.\citep{sun2024comparing} for more details of data generation process). In this setting, the treatment effect heterogeneity is relatively strong. The corresponding data and code used in the paper are available in a GitHub repository https://github.com/Novartis/WATCH.} 

\subsection{Analysis Planning}\label{sec:plan}
For analysis planning, it is essential to be familiarized with drug, disease and past (or possibly ongoing or planned) clinical trials, as well as engage with the project team and stakeholders. If possible, the planning of the analysis should take place prospectively, before data base lock. It is critical to understand the planned estimands (ICH E9 (R1) Addendum\citep{ICH2019}) and analyses, important prognostic factors (stratification and adjustment factors affecting the outcome variable independent of treatment) and the potential treatment effect modifiers (i.e. pre-specified exploratory subgroups). The scientific literature (same drug in other indications; different drugs with the same mechanism in the same indication) may reveal further prognostic variables or potential effect modifiers.

It is crucial to engage with analysis stakeholders and subject-matter experts to confirm that the aim of the analysis is to obtain an overview of how the treatment effect (that is the difference/ratio between two treatments) descriptively varies according to baseline covariates. Similar questions are often around prognostic modelling (risk modelling) of the outcome or cluster analysis in terms of baseline characteristics, which are out of scope for this workflow. Sometimes, the interest at outset might be to identify a subgroup with an enhanced, reduced or no treatment effect.  It is also important to understand from stakeholders the potential implications and decisions that might be made based on the planned analysis.

As a next step, it is necessary to align on the outcome variable(s), study/studies and baseline covariates to include. Regarding the outcome variable, \ks{one may be interested in a single outcome} variable or a small number of important outcome variables. In case multiple outcome variables are of interest, the analysis below would be repeated for each outcome variable. Findings are more convincing when replicated over several outcome variables (obviously replication of results over highly correlated outcomes is less convincing). Often it also makes sense to perform a sensitivity analysis for modifications of the original outcome variable of interest (e.g., for variables measured over time, using a different time-point, or averaging over time-points). 

The next step is to determine the scale of treatment effect. It is well known that the amount and even the existence of treatment effect heterogeneity depends on the treatment effect scale utilized \citep{kent2020a, yusuf2016interpreting}: Treatment effects may be homogeneous on one scale (e.g. relative ratio scale), but not on another scale (e.g. absolute difference scale): For example homogeneous treatment effects on a relative scale (e.g., risk ratio) will automatically be heterogenous on the difference scale (e.g., risk difference) as soon the outcome in the control group varies with patient covariates. The first scale to consider would typically be the one on which treatment effects for the endpoint is commonly presented (often a relative scale). Benefit-risk analyses, however, typically present treatment effects on a absolute difference scale, so that potentially multiple scales need to be considered.

If multiple studies are available, pooling can be considered. There are several pros and cons: A pooled analysis will have a larger sample size and thus potentially more information on treatment effect modifiers. On the other hand one must consider the similarities and differences of the studies to be pooled (method of data collection and definition of endpoint, dose, duration, population, how contemporaneous the studies are,...). For example a pooled analysis may be hard to interpret in a setting, where the there are differences in endpoint definition. When pooling is performed it is important to consider baseline population differences across trials (and arms) in particular if a treatment is only available in some of the studies (this then also needs to be appropriately reflected in the analysis). Finally, it typically makes sense to also repeat the analysis separately by study, \ks{or leave one study out to serve as an independent data set, and check the consistency of findings.}

A further crucial step is to determine the baseline variables to use in the analysis. These variables could be
\begin{itemize}
    \item factors of high plausibility, related to the mechanism of action of the drug, 
    \item factors that were effect modifiers for the same drug in other indications, or for different drugs with the same mechanism in the same indication,
    \item factors related to the severity or progression of the disease, or disease subtype,
    \item variables used as part of the design or pre-specified analysis, such as stratification variables, baseline values for continuous outcomes, or other adjustment factors,
    \item basic demographic information such as age, sex, race/ethnicity and weight,
    \item an established prognostic/risk score.
\end{itemize}
Similar as recommended in the EMA\citep{ema:2019} guideline, it is recommended to document the level of external, a-priori evidence for treatment effect modification. We propose to use the categories \emph{none}, \emph{low}, \emph{moderate}, \emph{high} for each variable, see Appendix \ref{sec:appex1} for details. We anticipate that the most common used category will be \emph{low}. For the (typically rare) categories \emph{moderate} and \emph{high}, a reference to the information source should be provided and documented, as well as the expected direction of the treatment effect. Ideally this takes place before any study or subgroup results were communicated. If the categorization happens afterwards an unbiased representation of a-priori evidence is no longer possible. We nevertheless recommend to categorize variables according to the study external evidence. Often this exercise would result in less than 10-15 covariates. It is an option to repeat the analysis based on a wider set of variables. Analysis results for this more wider set of covariates need to be critically assessed, as the number and plausibility of the included covariates has an important impact on the reliability of conclusions \citep{sun2024comparing}: When one searches for a needle in the haystack, adding hay won't usually help.

\subsection{Initial Data Analysis (IDA) and Analysis Dataset Creation}
This step includes two parts: initial data analysis (IDA) and analysis dataset creation. IDA describes the population distribution and covariate dependence. Based on that, one can further preprocess the original data (as discussed in Section \ref{sec:plan}) to generate an analysis dataset.\\

\noindent \textbf{Initial Data Analysis}\\
The aim of IDA is to explore the variables included in the analysis data sets to familiarize with the data and also to suggest transformation or omission of certain covariates. The main analysis question around treatment effect heterogeneity should not be approached in this step \citep{baillie2022ten}. Our IDA encompasses four steps, and a snapshot of plots that could result from this analysis are displayed in Figure \ref{fig:IDA}. 

We recommend to first investigate the distribution for each covariate via histogram or bar plots (Figure \ref{fig:uni_sum}). This step checks unusual patterns that may appear in the data distribution, such as imbalance between classes, skewed distributions, extreme observations etc. In the second step we produce summary visualizations of the covariates stratified by study and treatment (Figure \ref{fig:IDA_stra} shows an example). We recommend this step to check differences in the distributions among different factor levels, across studies or treatment arms in the same study. In the third step, we recommend to investigate missing values (Figure \ref{fig:IDA_miss}) and non-informative covariates. In terms of missing values, we recommend to check the percentage of missingness for each covariate, and the missing pattern across variables. Non-informative covariates can be categorical variables that are concentrated on essentially one factor level. In the final step we recommend analyzing the dependency among covariates by checking pairwise correlation and hierarchical clustering based on the correlations (Figure \ref{fig:IDA_dep}). This helps to identify identical or highly correlated variables and can also guide interpretation of results later. The selection of covariates to be kept usually needs to be discussed with the project team to make sure the most relevant and informative ones are retained.\\

\noindent \textbf{Analysis Dataset Creation}\\
After initial data analysis, certain changes need to be made to create the analysis dataset, such as variable transformations (for highly skewed covariates, or merging sparse categories for categorical variables), variable omissions (non-informative variables or nearly identical covariates) or imputation of missing baseline covariates. Concrete rules should be provided on how these data-set modifications are performed, limiting human input.

For imputing missing values in the covariates we suggest the imputation of baseline covariates to be independent of outcome or any other post-baseline variables. A variety of approaches are available for this purpose. We encourage users to run analysis with different imputation methods for a robustness check.\ks{Note that some methods don't require an imputation of missing baseline covariates (for example \cite{grf}), for those methods the step of missing covariate imputation can be omitted.}

For intercurrent events and missing data in the outcome variable we propose (for consistency) to follow the analytical strategy used in the main pre-specified clinical trial analyses for this endpoint, following the decided intercurrent event strategies and missing data handling approaches (see ICH E9 (R1) Addendum\citep{ICH2019}). In situations where these approaches are complex (e.g. various multiple imputation strategies used) subsequent analyses for predictive variable identification may become time-consuming and infeasible. In this case we suggest to use simpler approaches that are still in the spirit of the main estimand targeted in the clinical trial analyses. This may often imply using single imputation approaches.

\begin{figure}
\captionsetup{font=large}
    \centering
    \begin{subfigure}[b]{0.49\textwidth}
        \centering
        \includegraphics[width=\textwidth]{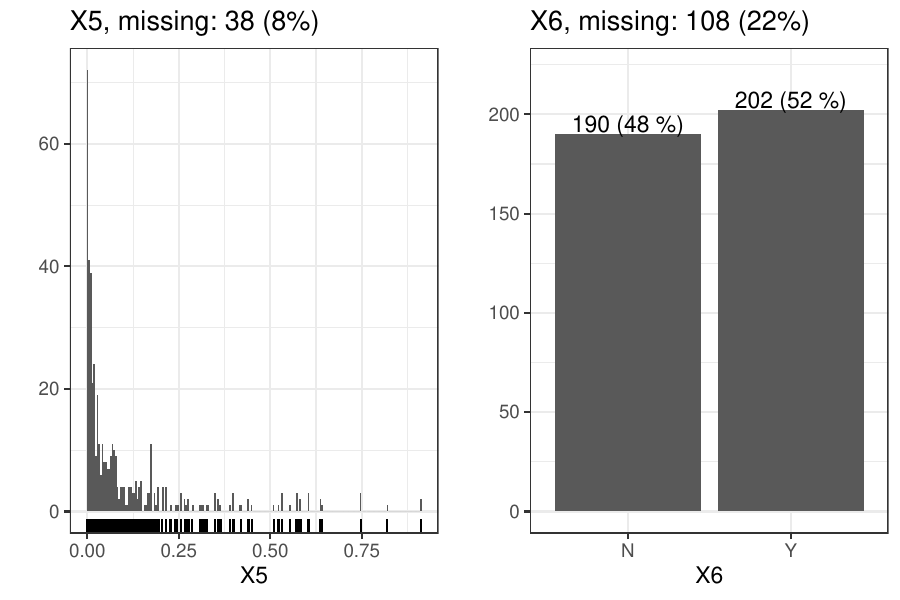}
        \caption{Univariate summaries}
        \label{fig:uni_sum}
    \end{subfigure}
    \begin{subfigure}[b]{0.49\textwidth}
        \centering
        \includegraphics[width=\textwidth]{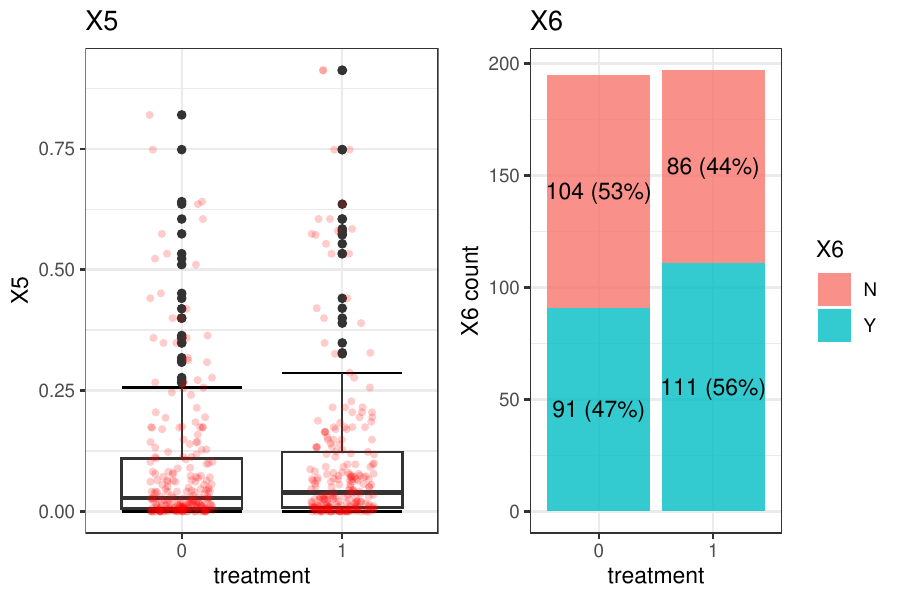}
        \caption{Variables stratified by further factors}
        \label{fig:IDA_stra}
    \end{subfigure}
    \begin{subfigure}[b]{0.49\textwidth}
        \centering
        \includegraphics[width=\textwidth]{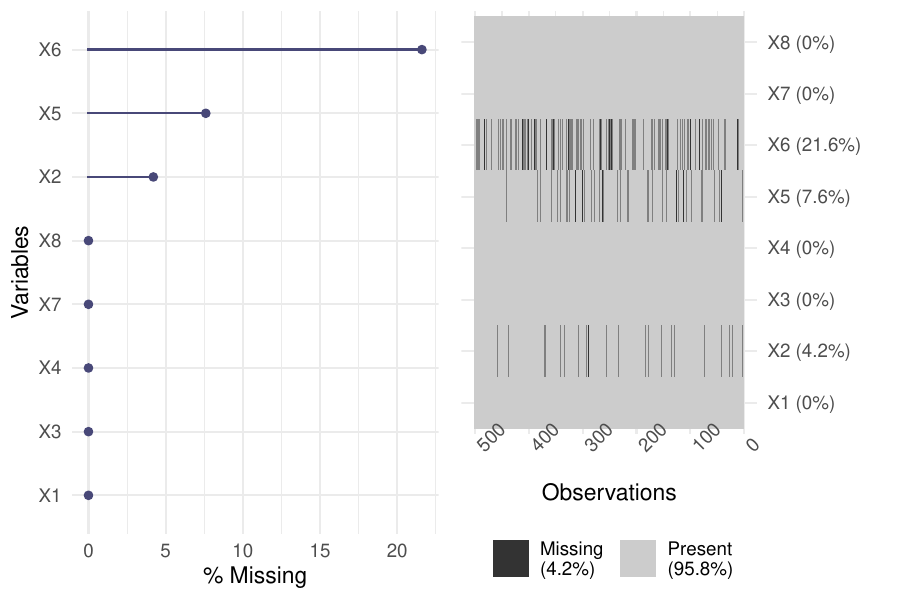}
        \caption{Missing values}
        \label{fig:IDA_miss}
    \end{subfigure}
    \begin{subfigure}[b]{0.49\textwidth}
        \centering
        \includegraphics[width=\textwidth]{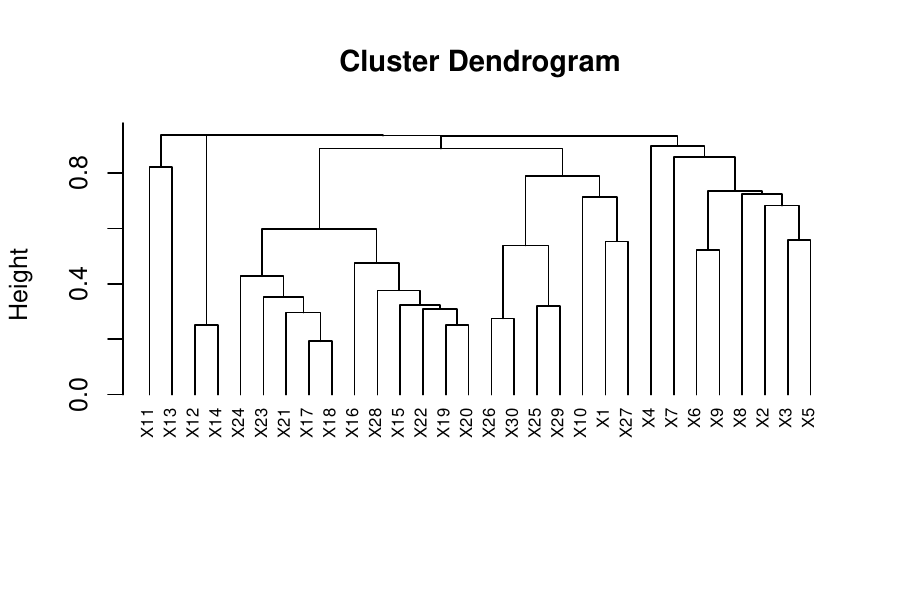}
        \caption{Dependence analysis}
        \label{fig:IDA_dep}
    \end{subfigure}
   
    \caption{Various visualisations generated after the Initial Data Analysis (IDA).\hspace{50cm}}
    \label{fig:IDA}
\end{figure}

\subsection{TEH Exploration}\label{sec:explore}
The exploration of TEH is the core analytical step of the proposed workflow. We propose to implement this by addressing three questions:
\begin{enumerate}[label={(Q\arabic*)}]
    \item How strong is the evidence against homogeneity?
    \item Which are the observed effect modifiers?
    \item How does the treatment effect change for the identified covariates? 
\end{enumerate}

Note that we recommend to structure the effort according to the three questions above; but different modelling approaches can be used for this purpose. An example would be to use (penalized) regression models. Based on this, for (Q1) \ks{an overall test could be performed} and for (Q2) a ranking based on the standardized model coefficients could be presented. Sun et al.\citep{sun2024comparing} performed simulations to compare several statistical methods on their performance to answer the above questions. For illustration, here we use an implementation on the double robust (DR) learner \citep{kennedy2023towards} combined with the conditional random forest \citep{hothorn2006unbiased}. The methodological details are not the main topic for this paper, and provided in Appendix \ref{sec:appex4}. The DR learner is attractive, as it provides a \emph{pseudo observation} that can be seen as an observation of the treatment effect for each individual. 
Based on these pseudo observations, one can then directly model the treatment effect in relation to baseline covariates (similar as proposed in the \emph{virtual twins} approach \citep{foster2011subgroup}). As a consequence, the covariates that interact with treatment can be directly identified from the model without modeling main effects. This set up can be used to address all three questions stated above. 
\newline

\noindent \textbf{Evidence Against Homogeneity}\\
Clinical trials are usually planned in a specific overall population with prior expectation of largely consistent treatment effects across that population (i.e. treatment effect homogeneity). Otherwise, the inclusion criteria of the trial and its primary analysis would have been different. 

It is thus a relevant question to ask: \ks{\emph{``How likely would it be to observe the actually observed heterogeneity under a data generation model that uses homogeneous treatment effects?''}}, while taking into account the number of variables utilized for assessing TEH. Answering this questions also provides useful information with respect to the reliability of possible observed heterogeneous treatment effects: If there is low evidence overall for treatment effect modification, identification of effect modifiers will be less reliable, and the estimation of their effects difficult and prone to random highs or random lows. A number of authors consider an \ks{overall test for the presence of TEH} as important: Interaction test are prominently mentioned in some of the workflows in the introduction \citep{Schandelmaier:2020, kent2020a, watson2020machine}. Harrell\citep{harrell:2023} proposes to use a global likelihood ratio test based on regression models, comparing a model with a treatment indicator and all baseline covariates as main effects, as well as interactions with treatment, to a model that includes only all baseline covariates and treatment as main effects. The PSI/EFSPI Working Group on Subgroup Analysis proposed the Standardised Effects Adjusted for Multiple Overlapping Subgroups (SEAMOS) plot. SEAMOS is intended to provide a graphical presentation of the treatment effects for all pre-specified subgroups to illustrate how extreme/surprising the observed results are if data were generated by a homogeneous treatment effect model\citep{dane2019subgroup}. Callegaro et al.\citep{callegaro2017testing} compare different options for \ks{interaction tests between treatment and covariates} in the context of clinical trials, when there are high-dimensional covariates.
Chernozhukov et al.\citep{chernozhukov2023} propose a machine learning approach to test for global effect modification, which is implemented in the grf R package \citep{grf} for the causal forest algorithm. \ks{Other methods to perform an overall tests for heterogeneous treatment effects, building on statistical and/or machine learning modelling are implemented in GenericML\cite{GenericML} and evalITR\cite{evalITR} R packages.} \ks{Finally, Lipkovich et al. \cite{lipkovich2024modern} provide a review of methods for testing for heterogeneity of treatment effect.} 

In practice, of course, some heterogeneity is always expected and exact homogeneity is almost never considered realistic. This, however, does not diminish the value of answering Q1, which quantifies how plausible it would be to observe the actually observed heterogeneity under a data generation model that uses homogeneous treatment effects.

The resulting p-value from such a \ks{overall assessment/test} should not be interpreted as a binary decision rule but on a continuum \ks{(i.e. considering the value rather than only dichotomizing according to a threshold)}. Trials are not planned for such tests and the statistical properties of such a decision rule would not be adequate for binary decision making \citep{Greenland2023}. A high p-value indicates compatibility, while a low p-value indicates contradiction of the data to a model of homogeneous treatment effects. The p-value should be interpreted as measuring the compatibility of the observed data with a model of homogeneous treatment effects \citep{cole2020}, see Appendix \ref{sec:appex2} for a proposed arbitrary verbal descriptions of the level of evidence against homogeneity. The \ks{overall test} p-value also provides a metric that can be compared across different situations (e.g., different trials, endpoints, different total number of covariates used). \ks{To avoid the risk of fishing expeditions, we recommend using a single method to derive the p-value, documenting it in the analysis plan. Additionally, simulations can be conducted upfront to decide which method to use.}


\ks{For the DR learner, testing against the homogeneity of treatment effect is equivalent to determining whether the pseudo observation of the treatment effect for each individual, denoted as $\widehat{\phi}(X)$, is independent of baseline covariates ,i.e. $\widehat{\phi}(X) \indep X.$ For this purpose we use an independence test as implemented in Hothorn et al. \citep{hothorn2006lego}; more details can be found in Appendix \ref{sec:appex4}. For the simulated data described earlier using this global heterogeneity test based on the DR learner, the p-value is 0.084, indicating a moderate evidence against homogeneity.}
\newline

\noindent \textbf{Effect modifiers}\\
Description of the baseline covariates that are most associated with the treatment effect in the observed data is essential. Variable importance can be used to determine the contribution of baseline variables towards treatment effects. However, there is no standard method for defining variable importance, and the implementation of variable importance calculation can vary \citep{WEI2015399}. One established approach is permutation importance, which involves permuting the covariates and measuring the increase in the prediction error of the model\citep{breiman2001random}. This importance measure is model agnostic and can be more easily implemented to compare the performance between different models. Variable importance derived from permutation importance represents multivariate importance, which measures the contribution of one covariate, with other covariates in the model. \ks{Another measure of variable importance that can be used is SHAP (SHapley Additive exPlanations) values, which provide insights into the contribution of each variable to the model’s predictions.\citep{Lundberg2017}}  

Other than exploring the variable importance for each covariate, one could also consider the variable importance based on paired variable contribution, see Molnar\citep{molnar2022} for more discussion on this topic. \ks{Note that the performance of many variable importance methods degrades in the presence of correlated variables, so extra care is needed when interpreting variable importance scores in such cases. If strongly correlated variables are included for justifiable reasons, robust methods should be used.\citep{Aas2021, debeer2020conditional}} \ks{Additionally, variable importance scores can be unstable and vary with small data perturbations or different random seeds. To quantify the stability of the findings, the analysis can be repeated on multiple bootstrap samples, with summaries of VI scores across these iterations presented, such as in the form of box plots. Additionally, the stability of predictive variable selection/ranking can be characterized using an approach such as described Nogueira et al.\cite{Nogueira2018}}

\ks{For the DR learner to derive importance scores we use the data $(X_i, \widehat{\phi}(X_i)) $ to build a conditional random forest model, where $\widehat{\phi}(X_i)$ are the the pseudo observation of the treatment effect for each individual.} Figure \ref{fig:importance} (a) presents the permutation importance based on conditional random forests \citep{hothorn2006unbiased} for the simulated data. Figure \ref{fig:importance} (b) presents the interaction variable importance based on partial dependence function \citep{greenwell2018simple} among the top 10 selected variables from Figure \ref{fig:importance} (a). This figure helps to narrow down the variables of interest for further exploration.  \ks{For the observed data $X_1$ is most strongly associated with the treatment effect (which we know is correct), while the other true treatment effect modifier $X_{14}$ is the 4th important variable from permutation importance, and the interaction between them is identified (not as the most important though) from the interaction importance from Figure \ref{fig:importance} (b).} \newline

\begin{figure}
    \centering
    \begin{subfigure}[t]{0.49\textwidth}
        \centering
        \includegraphics[width=\textwidth]{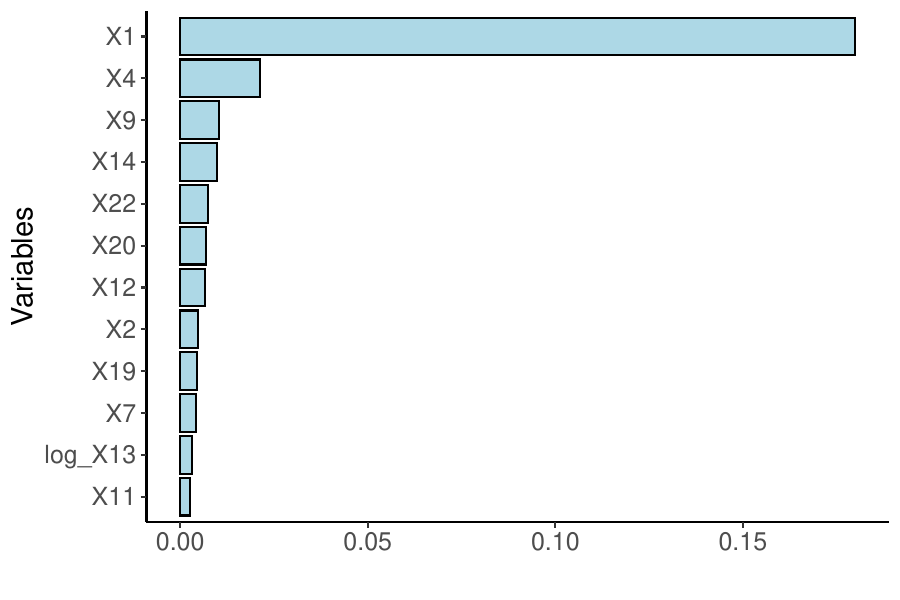}
        \caption{Permutation Variable importance}
    \end{subfigure}
    \begin{subfigure}[t]{0.49\textwidth}
        \centering
        \includegraphics[width=\textwidth]{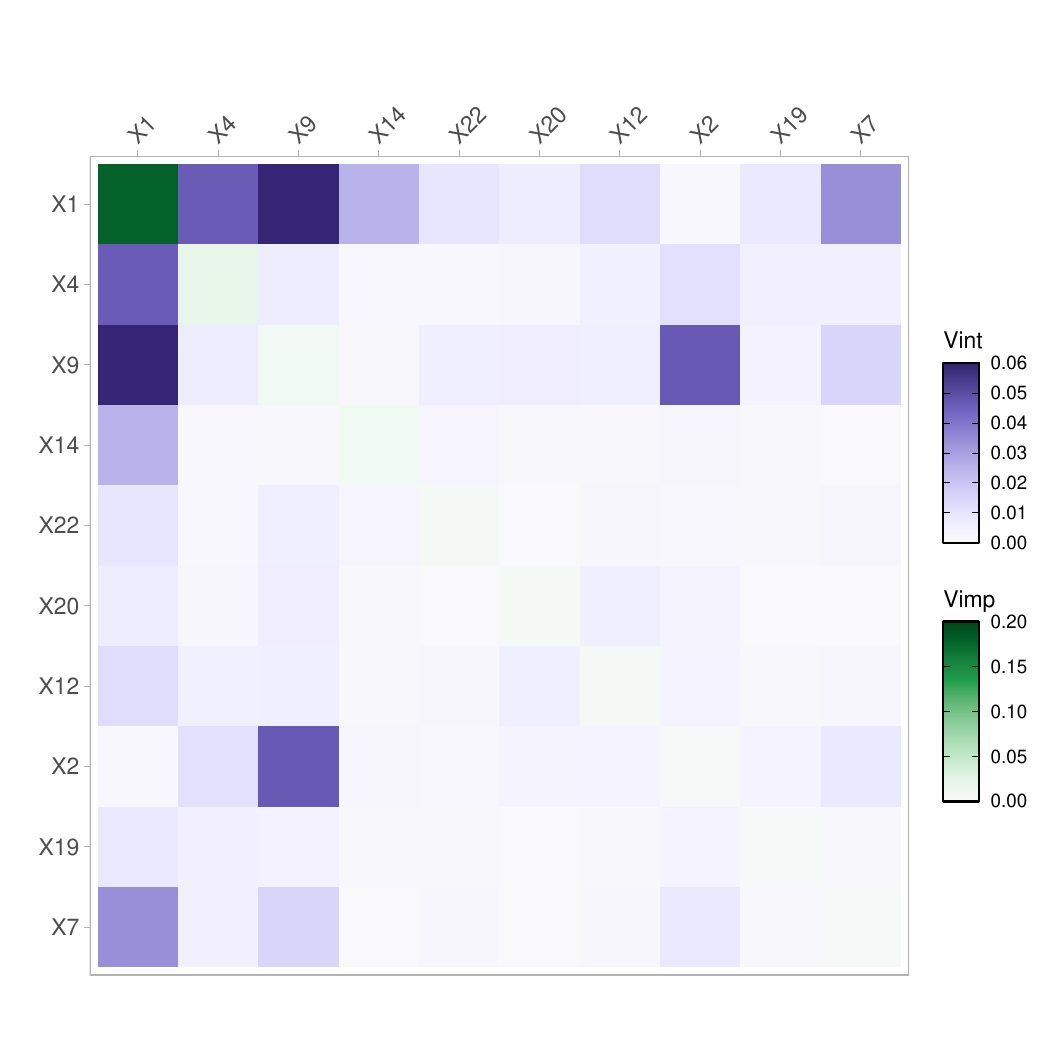}
        \caption{Interaction variable importance}
    \end{subfigure}
    \caption{Two different visualisations for the variable importance that captures the contribution of baseline variables towards treatment effects. Panel (a) is the permutation variable importance, Panel (b) includes permutation variable importance ``Vimp''(diagonal, which is equivalent to Panel (a)) as well as interaction variable importance ``Vint'' based on partial dependence function (off diagonal values).}
    \label{fig:importance}
\end{figure}

\noindent\textbf{Descriptive displays}\\
In order to investigate the covariates that appear high in the ranking, we recommend to present visualizations of how the treatment effect changes with respect to these covariates. This can be done in a univariate way, or in a bivariate way, i.e. checking pairs of variables. Furthermore, we propose to provide plots with the per arm response, and with the corresponding treatment effect. Please refer to the snapshot of these visualization plots in Figure \ref{fig:uni} and Figure \ref{fig:bi}. 

\begin{figure}
    \centering
    \begin{subfigure}[b]{0.49\textwidth}
        \centering
        \includegraphics[width=\textwidth]{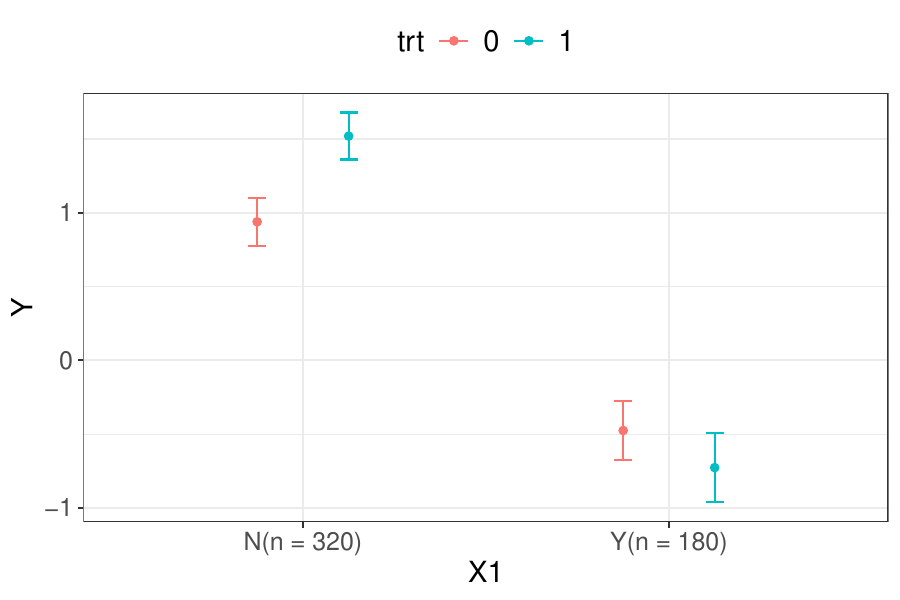}
        \caption{}
    \end{subfigure}
    \begin{subfigure}[b]{0.49\textwidth}
        \centering
        \includegraphics[width=\textwidth]{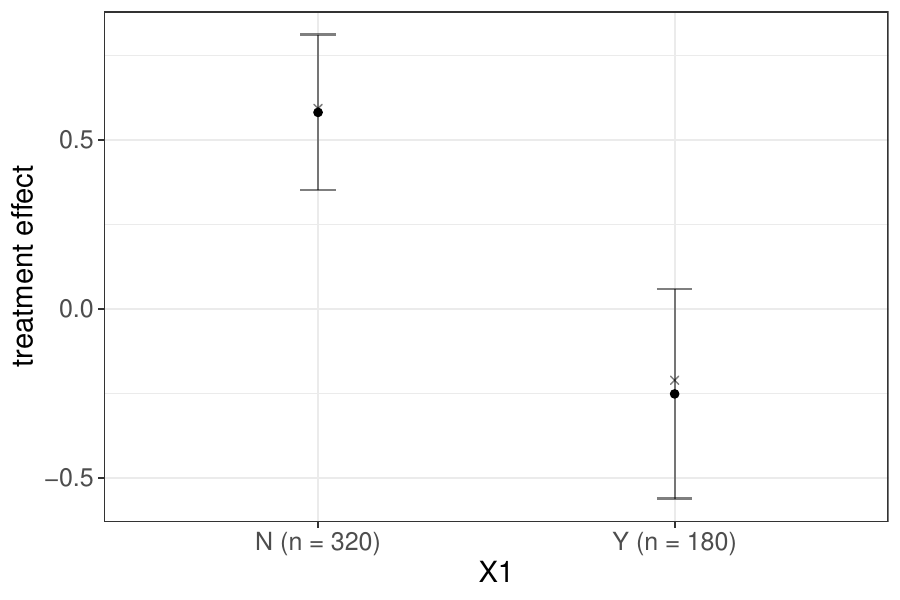}
        \caption{}
    \end{subfigure}
    \begin{subfigure}[b]{0.49\textwidth}
        \centering
        \includegraphics[width=\textwidth]{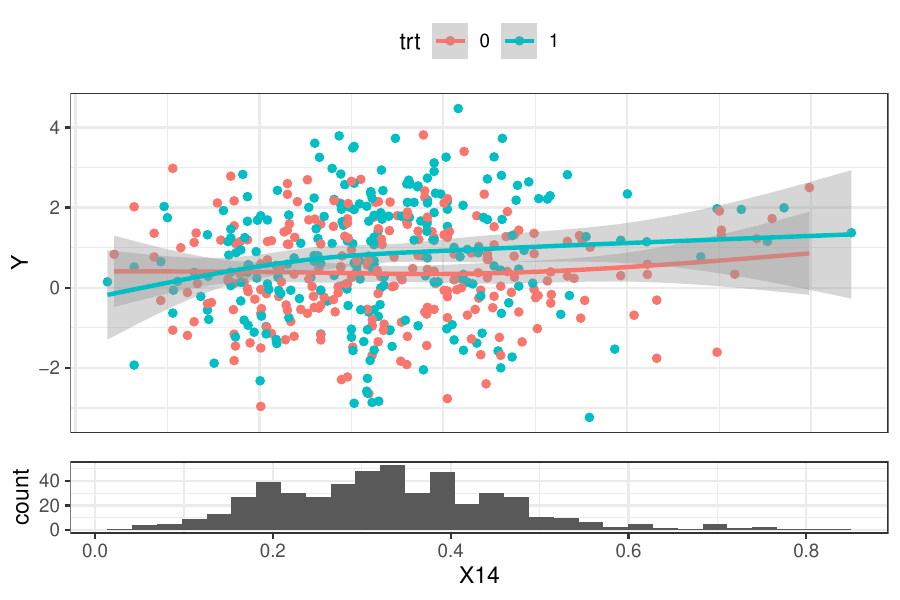}
        \caption{}
    \end{subfigure}
    \begin{subfigure}[b]{0.49\textwidth}
        \centering
        \includegraphics[width=\textwidth]{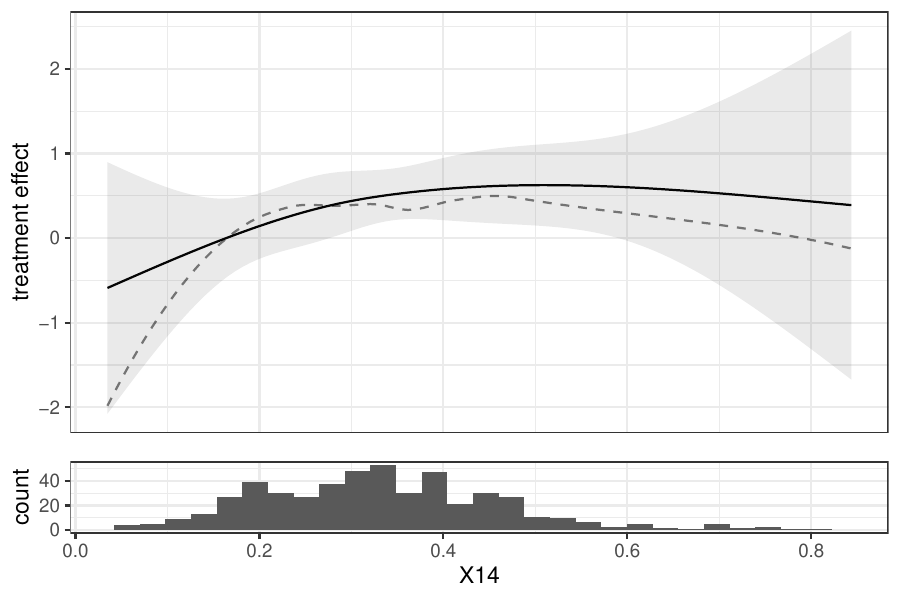}
        \caption{}
    \end{subfigure}
   
    \caption{Univariate display of outcome/treatment effect. Panel (a) 
    shows the outcome mean and associated confidence intervals using data within each group of $X_1$. In Panel (c) the  curve (with shaded pointwise 95\% confidence interval) is displayed in dependence on the covariate X27, estimated using a Generalized Linear Model (GLM) with a regression spline. Furthermore,  Panel (b) displays the unadjusted treatment effect with confidence intervals, as well as the average of the pseudo observations from the DR learner (stars). Panel (d) presents an unadjusted estimate of the treatment effect, along with confidence intervals derived from a regression spline. Additionally, it includes a smooth curve based on a local regression model applied to the pseudo observations from the DR learner.
    }
    \label{fig:uni}
\end{figure}

Figure \ref{fig:uni} illustrates the covariates (use $X_{1}$ and \ks{$X_{14}$} for demonstration) and treatment effects/outcomes. Panels (a) and (c) depict the impact of the covariates (categorical covariate in Panel (a) and continuous covariate in Panel (c)) on outcomes in different treatment arms (green for treated arm and red for placebo arm), while ignoring other covariates. 
The histogram of covariates (bottom subpanel from Panel (c)) and total number of patients in each group (from Panel (a)) provide information on the amount of data for interpretation. In addition to the outcome plots based on the covariates of interest, it is useful to provide corresponding plots for treatment effects as a comparison (Panels (b) and (d)). \ks{The star in Panel (b) and the dashed curve from Panel (d) are the estimation from the pseudo observation from DR learner.}

Aside from univariate plots, bivariate plots are also crucial in providing insights into the interaction effect of two covariates on treatment effects (see Figure \ref{fig:bi}). These types of plots enable to comprehend how the contribution of one covariate to treatment effects varies at different levels of the other covariate. Panel (a) of Figure \ref{fig:bi} demonstrates patient outcomes on treatment (green) and placebo (red) arms at different levels of combinations of $X_{1}$ and $X_{9}$ (which has the highest interaction importance, see Figure \ref{fig:importance} Panel (b)). For each subpanel, the point estimation (with confidence interval) is estimated in the same way as in Figure \ref{fig:uni}, stratified by the second variable. Similarly, Panel (b) demonstrates the relationship between two covariates and their contribution to treatment effects. Similarly continuous variables would be displayed as in Figure \ref{fig:uni} with regression splines, but stratified by the second variable. Because now the overall data are subset according to two variables, the conclusions get more variable and hence unreliable. Appendix \ref{sec:appex3} provides information on the correlation of the 10 baseline variables that have the largest variable importance.

As the overall interaction p-value here only indicated a moderate evidence against homogeneity the results in these plots would need to be cautiously interpreted. In fact, while the variables $X_1$ and $X_9$ interact strongly in the observed data on the treatment effect, this is not the case for the data generating model.

\begin{figure}
    \centering
    \begin{subfigure}[b]{0.49\textwidth}
        \centering
        \includegraphics[width=\textwidth]{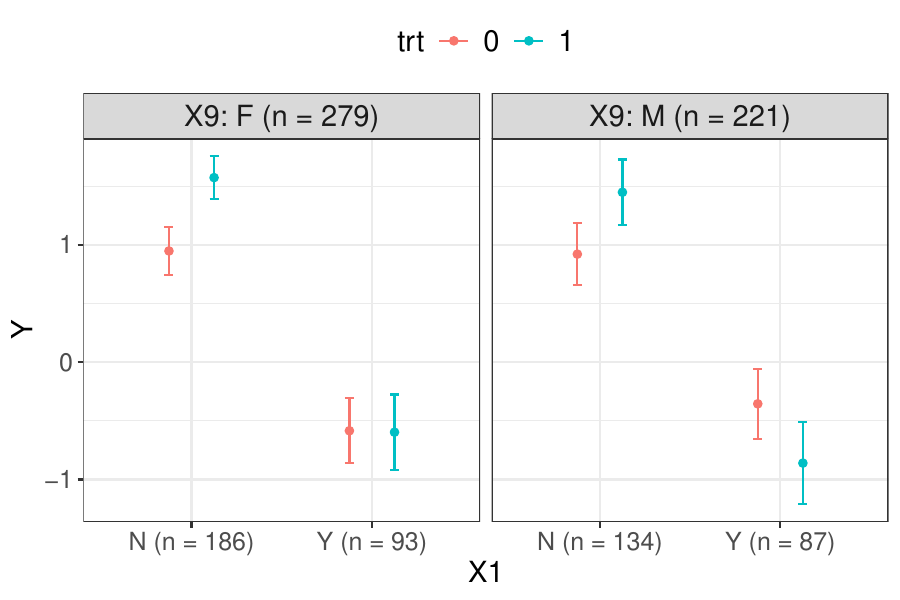}
        \caption{}
    \end{subfigure}
    \begin{subfigure}[b]{0.49\textwidth}
        \centering
        \includegraphics[width=\textwidth]{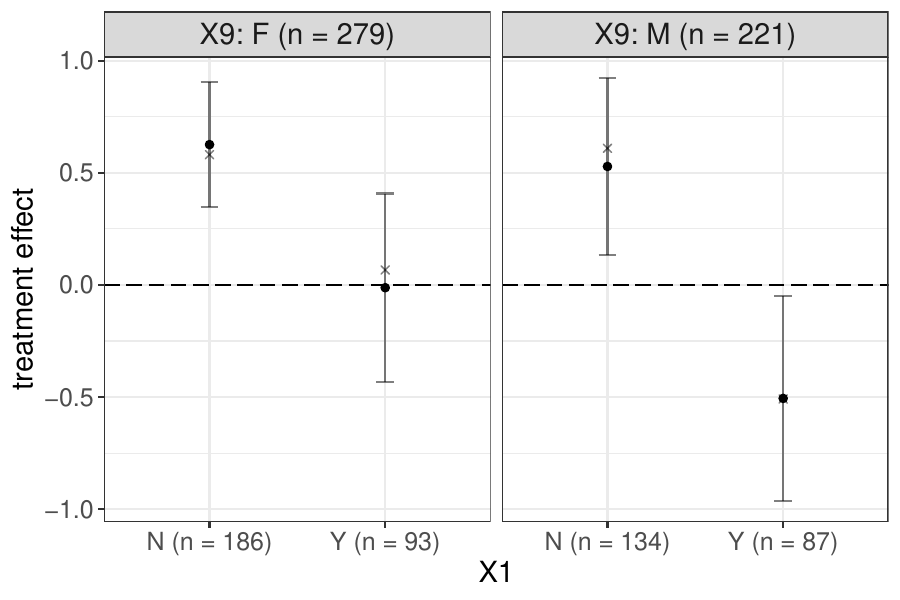}
        \caption{}
    \end{subfigure}
    \caption{Bivariate display of outcome/treatment effect. Panel (a) shows the outcome mean and associated confidence intervals using data within all possible groups defined by X1 and X9 jointly. Panel (b) displays the unadjusted treatment effect with confidence intervals, as well as the average of the pseudo observations from the DR learner (stars).}
    \label{fig:bi}
\end{figure}

It is a conscious choice that these plots show observed data and/or simple summaries  (means and regression splines) only and not results summarising a model fit (based on the conditional random forest in this case). An alternative display would be model-based partial dependence plots \citep{friedman2001greedy} that investigate the effect of a baseline variable on the outcome \emph{independent} of other variables, by effectively breaking correlations of the observed baseline variables for the display (using a form of standardization). This identifies which baseline variables are the drivers of heterogeneity for correlated variables. In our workflow this task is achieved by the model-based variable importance measure, but for display it is considered more appropriate to show the observed data and outcome means with the correlations as they appear in the data, which is more practically relevant (as correlations across baseline variables cannot be removed in the real world).
 
In addition to the bivariate plots, exhaustive subgroup plots \citep{muysers2020systematic} also provide a useful method for obtaining an overview of how treatment effects vary across subgroups; they have the appealing feature of putting the observed treatment effects into the context of many other subgroup treatment effects.


\subsection{Multidisciplinary Assessment}\label{sec:cfa}
As discussed in Section \ref{sec:intro}, the data-driven results obtained need to be assessed for credibility in light of existing external data and best scientific understanding. This is best achieved in a multidisciplinary discussion. 

The results from the workflow are descriptive and intended to be hypothesis-generating. Thus, strong inferential statements should be avoided when the results are communicated. \ks{The p-value from the overall assessment of evidence against homogeneity serves an overall diagnostic measure that guides on how much weight should be put on results that follow,} such as variable importance and treatment effect plots. The focus should be on a global overview instead of selective reporting of a single finding. To interpret the findings it is crucial to revisit the list of variables specified before start of the investigation as well as their external evidence and expected direction of the treatment effect, to assess whether the observed findings (and non-findings) are consistent with or different from existing external information. Data-driven findings that are not supported by a-priori, external evidence (categories \textit{none} and to some degree \textit{low} external evidence, see Appendix \ref{sec:appex1}) are of low credibility in particular if the global interaction p-value is large. On the other hand, strong findings with high external and a-priori evidence and correct pre-defined direction of treatment effect, should be considered very notable in particular when the global interaction p-value is smaller.  \ks{In the running example, we observed moderate evidence against homogeneity (Q1). Therefore, the recommendation on whether X1 can act as an effect modifier would depend on the a priori external evidence and the expected direction of the treatment effect.}

If further sensitivity analyses were conducted, then results could be mentioned and the focus should be on the similarities and differences in results compared to the main analysis. Sensitivity analyses could be with respect to using different endpoint, different time-point for the same endpoint, different covariate set, performing analyses \emph{by study} if multiple studies are involved, using different missing data imputation methods, using different treatment effect scales (e.g. relative versus absolute). If findings are replicated across several sensitivity analyses, they can be considered more robust compared to the situation where findings are not replicated. 
Analyses that are conducted in response to earlier analysis results and human input get more speculative in nature. For these analyses it is even more important to avoid definitive probabilistic inferential statements (as the model/data/variable selection process can no longer be quantified algorithmically).

Based on the information provided, the development team could consider different steps, depending on the drug development context: If the overall evidence and credibility of findings is low, this is useful strategic information, and the decision may be to nevertheless observe specific variables more closely in potential follow-up trials. In other cases where there is credibility for some of the findings, one might do further investigations to understand better the factors that could have caused the heterogeneity. In this case the investigation and analyses could be published externally and/or provide supportive information in health authority interactions. 

\ks{It is also possible that the team may be interested in identifying a specific subgroup as part of this discussion. This goes beyond this workflow, as it also depends on trade-offs that may need further stakeholder discussion (e.g.trade-off of subgroup size versus subgroup treatment effect; trade-off of subgroup efficacy versus subgroup safety). In this more inferential context, we suggest using methodologies such as shrinkage estimation of treatment effects to protect against random high selection bias, for example when making the choice on the treatment effect a new trial should be sample-sized for (see Thomas and Bornkamp\citep{thomas2017comparing} and Riehl et al.\citep{riehl2022shrinkage}).}

\section{Conclusions}
\label{sec:concl}
Understanding how treatment effects vary across patients can significantly influence sponsors' decisions related to drug development. These decisions are often based on internal exploratory analyses, which are challenging because clinical trials are not formally designed to investigate heterogeneity. This challenge underscores the importance of considering existing external data and the best scientific understanding. Therefore, we propose a workflow in this paper to describe and interpret treatment effect heterogeneity for trial sponsors in this non-inferential/exploratory setting \citep{tong2019statistical, muysers2020systematic}.

The core analytical part of the workflow consists of (i) an assessment of the global evidence against \ks{homogeneity} (according to the utilized covariates), (ii) providing a variable importance measure describing which baseline covariates are most associated with the treatment effect, and (iii) descriptive displays of how the treatment effect varies according to the important covariates. Assessment of the credibility of data-driven findings is facilitated due to an a-priori classification of the utilized baseline variables by external evidence and a final multidisciplinary assessment. Taking into account external information is consistent with the EMA guideline for subgroup analyses \citep{ema:2019}. 

In current practice, supportive, prespecified subgroup analyses are produced by default for important clinical trials and typically displayed in forest plots. In addition purely post-hoc exploratory analyses are performed (e.g. based on flexible statistical modeling and/or machine learning). While both approaches provide an overview of treatment effect heterogeneity, we believe the proposed workflow improves upon this:
\begin{itemize}
  \item It provides an overall assessment of evidence against homogeneity (rather than just univariate for each variable/subgroup in the forest plot), taking into account the number of utilized covariates (multiplicity).
  \item It is based on a multivariate assessment due to the chosen variable importance measures \citep{gregorutti2017correlation}, allowing \ks{us} to better understand the true drivers for heterogeneity, when there are correlated variables.
  \item It does not require to dichotomize continuous baseline covariates and more covariates can be assessed compared to the forest plot.  
  \item It allows \ks{us} to investigate on how two covariates interact on the treatment effect (which is usually not considered in a forest plot).
  \item It provides a systematic approach to summarize a-priori external evidence for treatment effect modification for each variable including a categorization of the corresponding evidence, which helps an assessment of credibility.
  \item Compared to more post-hoc exploratory analyses its results can be quickly be available after data-base lock, due to the pre-planning.
\end{itemize}  

While in subgroup analysis and treatment effect heterogeneity literature most of the works focus on developing and evaluating  analytical techniques, in our paper we wanted to emphasize that this is only one part in the work of a statistician or data scientist. The proposed WATCH workflow, that follows the PPDAC model, provides a systematic approach to explore treatment effect heterogeneity in the exploratory setting, taking into account external evidence and best scientific understanding. It has the potential to increase the quality of decision making around treatment effect heterogeneity and therefore suggest to plan for such an activity routinely at the time of design of any (sufficiently large) clinical trial. The importance of using structured problem-solving frameworks, such as PPDAC, to approach these exploratory endeavors has been highlighted in statistics literature\citep{wild1999statistical, spiegelhalter2019art}, and in biopharmaceutical research\citep{bretz2023role}. 

As we emphasized, our workflow focuses on the non-inferential/exploratory setting. For inferential subgroup identification a different workflow would be required, although some steps can be similar (e.g. analysis planning, deciding on the considered covariates and the existing a-priori/external evidence, the IDA and the analysis dataset creation). 

\ks{Finally, this paper illustrates the workflow using a running example implemented with the double robust (DR) learner combined with the conditional random forest. We would like to emphasize that the methods presented here are merely examples. There are numerous other approaches and modeling techniques that can be employed to implement this workflow, and we have reviewed some of them in this paper. However, a comprehensive review of all possible methods is beyond the scope of this paper and will be addressed in future work.}


\appendix

\section{Appendix}
\label{sec:appex}

\subsection{A-priori categorization of the level external evidence}
\label{sec:appex1}

We propose the following categorization of the level of external evidence for effect modification for specific variables.
\begin{itemize}
    \item The \emph{none} category should be used for variables that are included for reasons unrelated to plausibility or external evidence, but there is interest to investigate whether there is effect modification (depending on the context this could be variables such as age, sex or race). If a variable in this category comes up high in the importance ranking, this needs to be cautiously interpreted.
    \item The \emph{low} category is the default category, as it is expected that included variables are considered as \textit{plausible} effect modifiers, but most of the time without additional external evidence.
    \item The \emph{moderate} category is used in case there is external, a-priori evidence for treatment effect modification based on internal data or external presentation/publication, but it is not clear how relevant (e.g. different endpoint or indication but low interaction p-value after multiplicity adjustment) or reliable (e.g. moderately low interaction p-value and no multiplicity adjustment) this evidence is. 
    \item The \emph{high} category should only be used when there exists strong directly relevant evidence that a variable is a treatment effect modifier (e.g. low interaction p-value after multiplicity adjustment in previous study).  It is expected that \emph{high} will be very rarely applicable (otherwise, the underlying studies would have been designed in a different way). 
\end{itemize}
For the categories \emph{moderate} and \emph{high}, a reference to the information source should be provided and documented, as well as the expected direction of the treatment effect. 

\subsection{Verbal summary of level of evidence against homogeneity}
\label{sec:appex2}
Table \ref{tab:verbal} provides a verbal summary of level of evidence against homogeneity. The suggested scale for assessing evidence is based on the surprise value \citep{cole2020}, which is given by logarithm of base 2 of the p-value. It has a nice interpretation, since it represents the number of consecutive coin-flips with all heads up, under the hypothesis of a fair coin.
\begin{table}[h]
    \centering
    \begin{tabular}{|c|c|c|} \hline 
                     & surprise value & verbal summary of\\ 
      p-value        &  $-\log_2(p)$ & evidence against homogeneity\\ \hline      
      $[0.25,1]$       & $(0,2]$ & low \\
      $[0.063,0.25)$   & $(3,4]$ & moderate\\
      $[0.008,0.063)$  & $(5,7]$ & noteworthy\\
      $[0.001,0.008)$  & $(8,10]$ & strong\\
      $<0.001$       & $>10$ & very strong\\ \hline 
    \end{tabular}
    \caption{Suggested verbal summary of level of evidence against homogeneity.\hspace{50cm}}
    \label{tab:verbal}
\end{table}

\subsection{Correlation of baseline variables in simulated data}
\label{sec:appex3}
For the top 10 identified baseline covariates (based on variable importance), the density/bar plot for each variable, as well as the correlation (scatterplot/boxplot/bar plot) between each pair of variables are shown in Figure \ref{fig:correlation}.

\begin{figure}[!h]
  \begin{center}
    \includegraphics[width=0.95\textwidth]{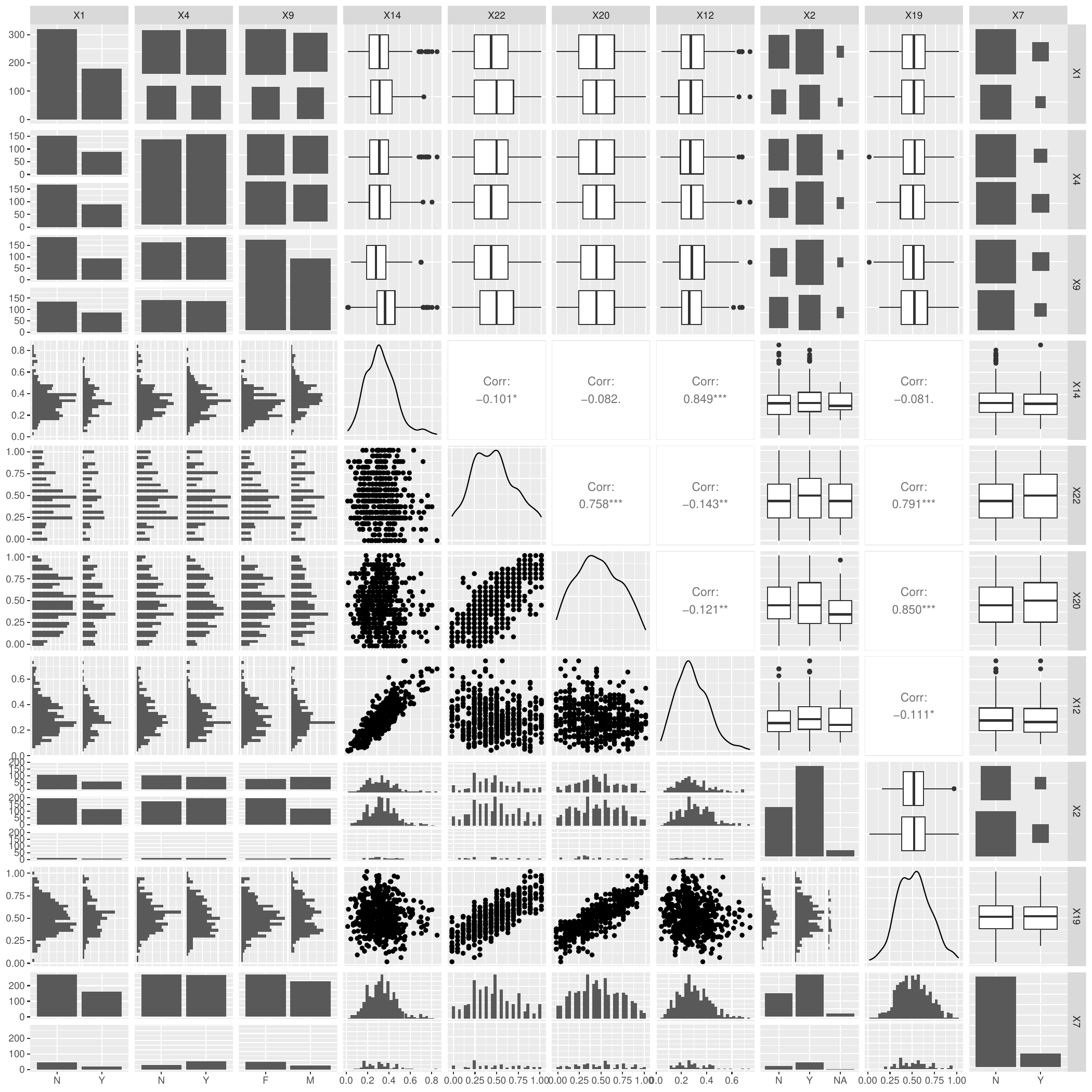}
  \end{center}
  \caption{Correlation of selected variables in the simulated data-set.\hspace{100cm}}
  \label{fig:correlation}
\end{figure}

\subsection{Details on the double-robust learner for example data}
\label{sec:appex4}
In this section we provide more details on the DR-learner\citep{kennedy2023towards} and how it was used in the presented example. Assume we have an i.i.d. sample of observations of $Z_i = (X_i, A_i, Y_i)$, where $X \in 	\mathbb{R}^p$ are covariates, $A \in \{0,1\}$ is a binary treatment, and  $Y \in \mathbb{R}$  an outcome of interest. We denote as $Y^1$  the potential outcome that would have been observed if the patient was receiving the treatment $A=1,$ while as $Y^0$ if the patient was receiving the treatment $A=0.$ 
The DR learner is a procedure  to estimate the conditional average treatment effect (CATE), $\mathbf{E} (Y^1 - Y^0 | X = x)$.  For doing this ``pseudo-outcomes'' are created for each patient, which can be seen as a noisy version of the CATE created. Algorithm \ref{alg:DR_learner} provides a detailed description of the construction of the pseudo outcomes for all patients and it is based on an algorithm described by Kennedy\citep{kennedy2023towards}. In Algorithm \ref{alg:DR_learner} we need to estimate three nuisance functions ($\widehat{\mu}_0, \widehat{\mu}_1, \widehat{\pi}$). To this end we can utilize  flexible regression models using machine learning. In our work we implement an ensemble of models using the \pkg{SuperLearner}, an algorithm that uses cross-validation to estimate the performance of multiple models, and then creates an optimal weighted average of those models (stacking) using the test data performance\citep{van2007super}. In our case we used penalized regression (LASSO), gradient boosting and random forests as the base learners.

\begin{algorithm}
\caption{DR learner}
Let $(D_1^n,...,D_K^n)$ denote K splits of n observations from the overall data based on a k-fold cross-validation. Let $D_{-k}$ denote the union of all cross-validation folds without the k-th fold $D_k^n$.
of $Z_i = (X_i, A_i, Y_i).$
\begin{description}
\item[Step 1.] Nuisance training:
\begin{description}
    \item[(a)] Construct estimates $\widehat{\pi}$ of the propensity scores $\pi(X)=\mathbf{P}(A=1|X)$ using $D_{-k}$.
    \item[(b)] Construct estimates $\widehat{\mu}_0, \widehat{\mu}_1$ of the regression functions ${\mu}_A(X)=\mathbf{E}(Y|X,A)$ using  $D_{-k}$.
\end{description}
\item[Step 2.] Pseudo-outcome: Construct the pseudo-outcome for all patients $i$ in $D_k^n$
\begin{align}
\widehat{\phi}(X_i) = \frac{A_i-\widehat{\pi}(X_i)}{\widehat{\pi}(X_i)\{1-\widehat{\pi}(X_i)\}} \{ Y_i - \widehat{\mu}_{A_i}(X_i) \} + \widehat{\mu}_1(X_i) - \widehat{\mu}_0(X_i)
\end{align}
Iterating steps 1 and 2 for all $K$ folds, This will provide a pseudo-outcome $\widehat{\phi}(X_i)$ for all patients.
\end{description}
\label{alg:DR_learner}
\end{algorithm}

Using pseudo-observations in a second step for inference for the treatment effect is convenient, because standard methods for modelling or regression can be used directly on the pseudo observations, without the need to model both main effects and interactions with treatment. As part of our workflow we don't directly need to estimate CATE, but are interested (i) in the evidence against homogeneity and (ii) to determine the treatment effect modifiers. To assess the evidence against homogeneity, we use the pseudo-observations to perform \ks{an overall test} against the homogeneity. We are interested in testing the following null hypothesis: $H_0 : \widehat{\phi}(X) \indep X.$ A number of methods can be used to perform that independence test, but in our implementation we use a test based on conditional inference procedures\citep{hothorn2006lego}. To answer the question on the important effect modifiers, we use again the pseudo-observations. To do that in a multi-variate fashion, we use the data $(X_i, \widehat{\phi}(X_i)) $ to build a conditional random forest model, and then rank the different covariates on their variable importance score. Because the clinical variables are of mixed data (continuous and categorical) it is important not to have any selection bias towards covariates with many possible splits. In our work implementation we use conditional inference trees\cite{hothorn2006unbiased}, an unbiased recursive partitioning method.

\section{Acknowledgments}
\label{sec:ack}
The authors would like to thank Frank Bretz, Malika Cremer, and Peter Quarg for their feedback and discussion on earlier versions of the manuscript. We also thank the EFSPI Treatment Effect Heterogeneity Special Interest Group (SIG) for their many discussions and feedback.

\bibliographystyle{abbrv}
\bibliography{bibl.bib}
\end{document}